\documentclass[conference]{IEEEtran}
\IEEEoverridecommandlockouts

\usepackage[ruled,linesnumbered]{algorithm2e}

\usepackage{url}
\usepackage{hyperref}
\hypersetup{
    colorlinks=true,
    linkcolor=blue,
    filecolor=magenta,      
    urlcolor=magenta,
    citecolor = magenta,
    pdftitle={An Example},
    pdfpagemode=FullScreen,
    }
\usepackage{xcolor}
\usepackage{pgfplots}
\usepackage{tikz}
\usepackage{comment}

\usepackage{filecontents}

\newcommand{\hlhref}[2]{\href{#1}{\textcolor{blue}{\underline{#2}}}}

\usepackage{threeparttable}

\usepackage{pifont} 
\usepackage{amssymb}
\usepackage{bbding} 
\usepackage{pifont}
\newcommand{\cmark}{\textcolor{green!80!black}{\ding{51}}}
\newcommand{\xmark}{\textcolor{red}{\ding{55}}}

\usepackage{makecell}
\usepackage{colortbl} 

\usepackage{fancyhdr} 
\usepackage{hyperref}

\usepackage{multirow}
\usepackage{cite}
\usepackage{amsmath,amssymb,amsfonts}

\usepackage{textcomp}
\usepackage{xcolor}

\usepackage{subfigure}
\usepackage{pgfplots}
\usepackage{filecontents}

\usepackage{graphicx,subfigure}

\definecolor{bblue}{HTML}{4F81BD}
\definecolor{rred}{HTML}{C0504D}
\definecolor{ggreen}{HTML}{9BBB59}
\definecolor{ppurple}{HTML}{9F4C7C}

\usepackage{listings}

\usepackage{blindtext}
\usepackage{etoolbox,xstring,mfirstuc,textcase}
\usepackage{booktabs}
\usepackage{pgfplots}

\usepackage{tabularx,booktabs,makecell,multirow, caption}
\usepackage{enumitem} 

\newcolumntype{L}{>{\arraybackslash}X}

\usepackage{xcolor}
\usepackage{tikz}
\usepackage{pgfplots}

\definecolor{findOptimalPartition}{HTML}{D7191C}
\definecolor{storeClusterComponent}{HTML}{FDAE61}
\definecolor{dbscan}{HTML}{ABDDA4}
\definecolor{constructCluster}{HTML}{2B83BA}

\usepackage[utf8]{inputenc}
\usepackage{listings}
\begin{document}
\title{Account Abstraction, Analysed}



\author{Qin Wang\IEEEauthorrefmark{1}, Shiping Chen\IEEEauthorrefmark{1}
\\
\IEEEauthorrefmark{1}\textit{CSIRO Data61, Australia}
}

\maketitle

\begin{abstract}

Ethereum recently unveiled its upcoming roadmap's \textit{Splurge} phase, highlighting the integration of EIP-\hlhref{https://eips.ethereum.org/EIPS/eip-3074}{4337} as a foundational standard for account abstraction (AA). AA aims to enhance user accessibility and facilitate the expansion of functionalities. Anticipatedly, the deployment of AA is poised to attract a broad spectrum of new users and ignite further innovation in DApps. In this paper, we elucidate the underlying operating mechanisms of this new concept, as well as provide a review of concurrent advancements in accounts, wallets, and standards related to its development. We step further by conducting a preliminary security evaluation to qualitatively assess the extent of security enhancements achieved through AA updates. 
\end{abstract}

\smallskip

\section{Introduction}
\label{sec-intro}

\begin{figure}[!htb]
    \centering
	\includegraphics[width=0.99\linewidth]{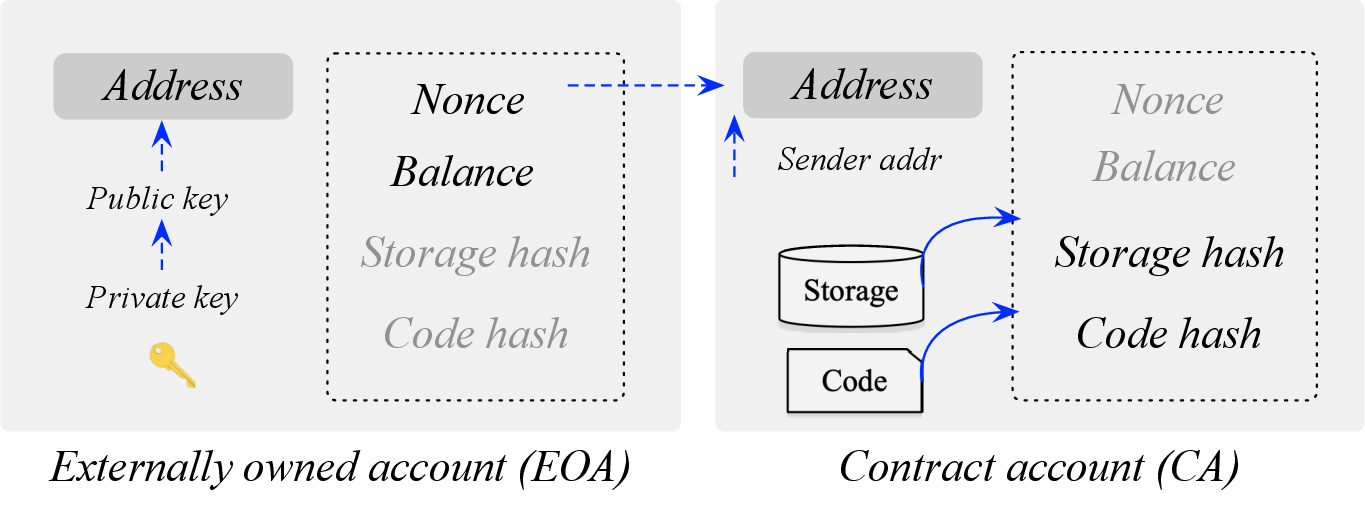}
	\caption{Ethereum accounts~\cite{evm}}\label{fig_account}
 \vspace{-0.4em}
\end{figure}

Accounts within the Ethereum ecosystem \cite{wood2014ethereum} serve as the bedrock for asset querying, storage, and transactions, constituting a pivotal element of its infrastructure. However, the present account design poses challenges for Web2 users due to its complicated design. Ethereum classifies accounts into two types~\cite{account}, namely, \textit{externally owned accounts} (EOA) and \textit{contract accounts} (CA)\footnote{Across several documents, the term \textit{smart contract account} has been used interchangeably to denote the same concept within the context of this paper.}. EOA is controlled by private keys held by its creators or users. Contract accounts are controlled by codes without the involvement of private keys (Fig.\ref{fig_account}). The reliance on private keys makes these two accounts act in different roles: EOA can prove the validity of a transaction and trigger the state transition in CAs but with limitations:

\begin{itemize}
  \item \textit{Expense}: The functioning of the contract wallet necessitates initiation by an EOA, essentially entailing a contract invocation. Every transaction within this process incurs an extra 21,000 Gas cost, which includes charges for ECDSA signature verification, Nonce value verification, and ensuring adequate account balance.
  \item \textit{Elevated barrier}: EOAs must possess a substantial ether balance to cover Gas expenses (involving management of two separate accounts), or alternatively depend on a Relayer to manage Gas payments, potentially introducing centralization concerns.
  \item \textit{Flunction}: In addition to the fee in ether (ETH), users are consequently required to hold ether, which exposes them to the potential volatility of its price.
  \item \textit{User perception}: From a user's perspective, grasping the nuances of gas price, gas limit, and transaction congestion is far from straightforward.
\end{itemize}

The effort to integrate two distinct types of accounts while preserving their core functionalities has been a longstanding subject of deliberation within Ethereum communities. Two primary technical pathways have come to the forefront.

\begin{itemize}
    \item EOA delegates control to smart contracts, in which the contract logic can inplement the core functionalities from EOA transactions such as gas payment (e.g., EIP\footnote{Ethereum Request for Comment (ERC) is one type of Ethereum Improvement Proposals (EIP) that requires on-chain agreements. Throughout this paper, we use EIP to refer to all standards for the sake of simplicity.}-\hlhref{https://eips.ethereum.org/EIPS/eip-3074}{3074}).
    \item EOA accounts are designed to be armed with several key functionalities from smart contracts (EIP-\hlhref{https://eips.ethereum.org/EIPS/eip-4337}{4337}).
\end{itemize}

The concept of \textit{account abstraction} (AA), indexed by EIP-\hlhref{https://eips.ethereum.org/EIPS/eip-4337}{4337}~\cite{erc4337}, aligns with the second approach, aiming to bestow EOA with the programmable functionality akin to CAs. The incorporation of EIP-4337 into the present roadmap~\cite{ethereumaa} signifies Ethereum's definitive stance in these dual directions. By adopting AA updates, Ethereum sidesteps the potential hurdle of transitioning existing users. This is attributed to its implementation on the application layers rather than the consensus layer, which inherently offers robust backward compatibility capabilities. Ethereum transactions are required to emanate from EOAs, which are typically accessed through various wallets like MetaMask, Phantom, and Rainbow. The account abstraction approach ensures the retention of current users while empowering them with an effortless means of engaging with smart contracts. Beyond that, the functionalities elucidated in EIP-4337 can be further applied across Ethereum-compatible blockchains, including platforms such as BNB Chain (formerly BSC), Polygon, Avalanche, Optimism, Arbitrum, and Base (also refer to Sec.\ref{subsec-supports}).

As of August 2023, a sequence of indicators\footnote{Dune Analytics \url{https://dune.com/niftytable}} highlights the escalating attention garnered by this impactful concept. The cumulative count of active accounts has reached 739,295, and the aggregate count of successfully executed $\mathsf{UserOperation}$s (detailed explanations refer to Sec.\ref{sec-aa}) has reached 1,307,197. In addition, the total number of bundled transactions has amounted to 1,197,871.
Besides, in August, a remarkable upswing is observed across various market indicators compared to previous months (starting from March 2023, which marks the launch of EIP-4337): The count of monthly active EIP-4337 smart accounts has surged to an impressive 420k (inclusive of all platforms). The earnings from monthly revenues and UserOperation fees have surpassed 24k and 220k respectively. Notably, the monthly tally of successfully executed UserOperations stands at 710k, while the gas expenses covered by paymasters have exceeded \$360k USD.

\smallskip
\noindent\textbf{Our attempts.} Given its nascent stage, only a limited number of studies have delved into this emerging concept. We are aligned with the growing momentum surrounding account abstraction and further expand the boundaries of its research landscape. In particular,

\begin{itemize}
\item \textit{\textcolor{teal}{Concept refinement}} (Sec.\ref{sec-account}\&\ref{sec-aa}). We conduct a thorough exploration of diverse resources, including academic literature, blogs, forum discussions, and Git repositories. Drawing from such analyses, we offer a coherent and succinct elucidation of the fundamental concepts and operational mechanisms of account abstraction.

\item  \textit{\textcolor{teal}{Security framework}} (Sec.\ref{sec-security}). We further delve into the security risks that may exist in AA. Delving into historical contract and account-related vulnerabilities, we construct a framework (Tab.\ref{tab:vulner}) to engage in a proper discussion about the security implications of this account paradigm.

\item  \textit{\textcolor{teal}{Further discussion}} (Sec.\ref{sec-discu}). Additionally, we provide more discussion about the principle of account abstraction designs and potential opportunities inspired.
\end{itemize}

\smallskip
\noindent\textbf{Security results.} Our security discussion (majorly Sec.\ref{sec-security}) led to the following insights. It is evident that integrating AA into current account systems can alleviate a multitude of vulnerabilities in both contract usage and block creation domains. However, AA's effectiveness remains limited when addressing intrinsic aspects of solidity language and design intricacies, such as structural elements, configuration constraints, and verification mechanisms. Upon further analysis, we have identified that AA's strength in fortifying security arises from its decoupling of key account abstraction components. By reallocating functions previously encompassed within accounts, AA can protect against numerous vulnerabilities in Ethereum's application layer: critical functions like gas payment mechanisms are seamlessly transferred to the paymaster, avoiding vulnerabilities related to gas fees; the intricate process of transaction packaging now benefits from the management of bundler; and complex trading and swapping operations, previously reliant on CEX or DEX, are elegantly executed within the same transaction. The decoupling process has reduced the threat that was heavily dependent on contract functions. thereby increasing the overall security quotient.

\smallskip
\noindent\textbf{Available resources.} The Ethereum official documentation \cite{ethereumaa} has introduced formal documentation detailing its account abstraction concept and roadmap that is accompanied by a suite of relevant standards, encompassing EIP-\hlhref{https://eips.ethereum.org/EIPS/eip-2771}{2771}, \hlhref{https://eips.ethereum.org/EIPS/eip-4337}{4337}, \hlhref{https://eips.ethereum.org/EIPS/eip-2938}{2938}, and \hlhref{https://eips.ethereum.org/EIPS/eip-3047}{3047}. Notably, EIP-4337 encapsulates the central concept driving this transition. In a study by Singh et al. \cite{singh2023account}, an initial exploration into the Ethereum account abstraction is provided, outlining its distinctive features and functional methodologies. Binance Research \cite{binanceaa} has released an insightful analysis that delves into recent market trends and noteworthy advancements within this area. Based on these contributions, a range of media sources \cite{metamaskaa,Alchemyaa,coinmarketcapaa,Hackernoonaa} have embarked on elucidating this intricate concept, offering surrounding insights and explanations. Besides, several studies have initiated the development of on-top solutions aimed at addressing various challenges within the Web3 ecosystem~\cite{wang2022exploring}, such as address incompatibility~\cite{park2023beyond}.

\section{Ethereum Accounts}
\label{sec-account}

\subsection{Recall Ethereum Account}

\noindent\textbf{EOA.} EOAs serve the purpose of storing and transferring ether, and ERC-20 tokens. EOAs are generated from a public key, which is a 20-byte hexadecimal identifier (e.g., 0xF57D1D6b84db4053cE452B35B7DB77878dCbdc65). These accounts are managed by a private key, which includes the EOA's password or seed phrase, held exclusively by the account's owner. Transactions involving EOAs do not rely on code or smart contract logic for their validity. As long as the private key remains known, the account's owner possesses the capability to execute transactions. The transaction's verification is contingent upon the user's signature and nonce.

\smallskip
\noindent\textbf{Contract account.} CA is a type of account that executes operations based on its pre-programmed logic, thereby enabling the creation of decentralized applications (DApps) and facilitating various functionalities within the Ethereum network. CA autonomously executes code in response to transactions, potentially modifying the contract's state, and exchanging ether or tokens.  CA accounts are assigned unique addresses akin to individual identification numbers, which permit interaction with other accounts. Once a CA is deployed, its code and state become immutable, contributing to transparency and ensuring the integrity of transactions. They also offer storage capabilities for manipulating data on the blockchain (Fig.\ref{fig_account}). These accounts can interact with EOAs and generate events that facilitate communication with other accounts.

\smallskip
\noindent\textbf{Two types of transaction.} These different accounts give rise to two distinct practical types of transactions for communication: \textit{contract creation} and \textit{message call}. Contract creation involves the generation of a new smart contract, with the transaction carrying an initialization code segment to define the new contract's properties. This process results in the assignment of a unique address to the newly created contract, which includes both its code and storage within the corresponding account state. Conversely, a message call signifies the modification of a smart contract's state. In this case, the transaction includes input data to update the contract's internal data. A message call does not create a new contract in the world state; instead, it alters the existing contract's state.

\subsection{Challenges in Account Design}

Coming with private keys, EOA has many fundamental functions including claiming the \textit{ownership} of the account and \textit{signing} the permissions of transactions. However, this may present a number of significant concerns:

\begin{itemize}

    \item \textit{Risk of private key loss.} Users who lose their private keys (due to loss or hacking) would face the irreversible loss of all their assets.

    \item \textit{Restricted signature options.} The native protocol exclusively supports ECDSA signature and verification algorithms for transaction validation.

    \item \textit{Single signer authority.} The absence of inherent multi-signature capability (which can only be achieved through smart contracts) means that a single signature is all that's required to execute any operation.

\end{itemize}

\subsection{Ethereum Roadmap}
Account abstraction is a crucial functional improvement outlined in the sixth phase of Ethereum's roadmap (Tab.\ref{tab:roadmap}). This upgrade involves a series of smaller refinements and adjustments aimed at ensuring seamless network operations subsequent to the implementation of other upgrades.

\begin{table}[!hbt]
\caption{Ethereum's roadmap}\label{tab:roadmap}
\vspace{-0.15in}
\begin{center}
\resizebox{1\linewidth}{!}{
\begin{tabular}{c|r} 
\cellcolor{gray!10}\textit{\text{Merge}}     &  \cellcolor{gray!10}Move to proof-of-stake consensus\\ 
\cellcolor{gray!10}\textit{\text{Surge}}     & \cellcolor{gray!10}Rollup-centric scaling to 100,000+ transactions per second \\
\cellcolor{gray!10}\textit{\text{Scourge}}     &  \cellcolor{gray!10}Avoid centralization and other protocol risks from MEV \\
\cellcolor{gray!10}\textit{\text{Verge}}     & \cellcolor{gray!10} Verifying blocks to become super easy \\ 
\cellcolor{gray!10}\textit{\text{Purge}} & \cellcolor{gray!10} Simplify the protocol and reduce the costs of running nodes  \\ 
\cellcolor{gray!10}\textit{\text{Splurge}} & \cellcolor{gray!10}Fix everything else (\textcolor{blue}{account abstraction} included) \\ 
\end{tabular}
}\end{center}
\end{table}

\section{Account Abstraction}
\label{sec-aa}

\subsection{Previous Attempts}

A series of continuous efforts have been made to make existing accounts extensible with more functions. We list a series related token standards in Tab.\ref{tab:erc}.

\begin{table}[!hbt]
\caption{AA-related Ethereum token standards}\label{tab:erc}
\renewcommand\arraystretch{1.1}
\begin{center}
\resizebox{\linewidth}{!}{
\begin{tabular}{l|ccc} 
        \cellcolor{gray!10}\textbf{EIP-} &\cellcolor{gray!10} \textbf{Time} & \cellcolor{gray!10}\textbf{Status} & \cellcolor{gray!10}\rotatebox{0}{\textbf{Key}}   \\
        \cmidrule{1-3}
        \cellcolor{gray!10}\hlhref{https://eips.ethereum.org/EIPS/eip-101}{101}
        & \cellcolor{gray!10} 2015 &\cellcolor{gray!10} Stagnant & \cellcolor{gray!10}Serenity currency and crypto abstraction \\ 
        \cellcolor{gray!10}\hlhref{https://eips.ethereum.org/EIPS/eip-86}{86} & \cellcolor{gray!10} 2017 &\cellcolor{gray!10} Stagnant & \cellcolor{gray!10} Abstraction of transaction origin and signature \\ 
        \cellcolor{gray!10}\hlhref{https://github.com/ethereum/EIPs/issues/859}{859} & \cellcolor{gray!10} 2018 & \cellcolor{gray!10} Stagnant & \cellcolor{gray!10} Account abstraction for main chain \\ 
        \cellcolor{gray!10}\hlhref{https://eips.ethereum.org/EIPS/eip-2718}{2718} & \cellcolor{gray!10} 2020 & \cellcolor{gray!10} n/a & \cellcolor{gray!10}  Typed transaction envelope \\ 
        \cellcolor{gray!10}\hlhref{https://eips.ethereum.org/EIPS/eip-2938}{2938} & \cellcolor{gray!10} 2020 & \cellcolor{gray!10} Stagnant & \cellcolor{gray!10}  Account abstraction \\ 
        \cellcolor{gray!10}\hlhref{https://eips.ethereum.org/EIPS/eip-3607}{3607} & \cellcolor{gray!10} 2021 & \cellcolor{gray!10} n/a & \cellcolor{gray!10} Reject transactions from senders with deployed code \\ 
         \cellcolor{gray!10}\hlhref{https://eips.ethereum.org/EIPS/eip-5003}{5003} &\cellcolor{gray!10}  2022 & \cellcolor{gray!10} Stagnant & \cellcolor{gray!10} Insert Code into EOAs with AUTHUSURP \\
        \cellcolor{gray!10}\hlhref{https://eips.ethereum.org/EIPS/eip-5189}{5189} & \cellcolor{gray!10} 2022 & \cellcolor{gray!10} Stagnant & \cellcolor{gray!10} Account abstraction via endorsed operations  \\ 
        
        \cmidrule{1-2}
        
        \cellcolor{gray!10}\hlhref{https://eips.ethereum.org/EIPS/eip-2771}{2771} & \cellcolor{gray!10} 2020 &\cellcolor{gray!10}  Final & \cellcolor{gray!10} Secure Protocol for native meta transactions \\ 
        \cellcolor{gray!10}\hlhref{https://eips.ethereum.org/EIPS/eip-3074}{3074} & \cellcolor{gray!10} 2020 &\cellcolor{gray!10}  Review & \cellcolor{gray!10} Allow EOA to delegate control to a contract  \\ 
        \cellcolor{gray!10}\hlhref{https://eips.ethereum.org/EIPS/eip-6900}{6900} & \cellcolor{gray!10} 2021 & \cellcolor{gray!10} Draft & \cellcolor{gray!10} Modular Smart Contract Accounts and Plugins \\ 
        \cellcolor{gray!10}\hlhref{https://eips.ethereum.org/EIPS/eip-4337}{4337} & \cellcolor{gray!10} 2021 & \cellcolor{gray!10} Draft &  \cellcolor{gray!10}  Account abstraction using alt mempool  \\ 
         \cellcolor{gray!10}\hlhref{https://eips.ethereum.org/EIPS/eip-6551}{6551} & \cellcolor{gray!10} 2023 & \cellcolor{gray!10} Review & \cellcolor{gray!10} Non-fungible token bound accounts 
        \\ 
\end{tabular}
}
\end{center}
\vspace{-0.2in}
\end{table}

In Ethereum's early phases, attempts were made to distinguish between EOA and CA by introducing various new transaction types. EIP-86/208 leverage the differentiation between two accounts to design customizable and collision-resistant contract addresses. This, in turn, led to the implementation of EIP-1014 and EIP-2470.
EIP-859 introduced transaction-initiated contract deployment. It allowed for on-the-spot deployment of contract addresses if none existed, forming the basis for the fundamental functionality of EIP-4337. EIP-2718 ushered in compatibility for future Ethereum iterations with any newly suggested transaction types. EIP-293 systematically cataloged various advantages of contract accounts, encompassing recovery, key rotation, customized identity verification algorithms, and meta-transactions, which collectively entrenched the role of contract accounts.

However, these efforts resulted in unwieldy complexity. Altering transaction types requires concurrent modifications to the underlying signature verification algorithms. This encompasses considerations like miner acceptance of new types, ensuring incentives are not lower than regular transactions to encourage verification, and addressing concerns about account address management and conflicts. Unfortunately, this approach lacked both backward and forward compatibility. Ultimately, these attempts have coalesced around two primary directions as stated in our Sec.\ref{sec-intro}.

\subsection{Account Abstraction}

Account abstraction introduces a suite of functions designed to personalize fundamental elements that enable smart contract functionalities. These customizations encompass various aspects, including user operations, fee payment methods, and transaction packing mechanisms.

\smallskip
\noindent\textcolor{teal}{\textit{Overview.}} 
The workflow is summarized as follows: \ding{172}~Users initiate interactions with the frontend abstraction (also referred to as application) layer, where their actions are translated into underlying transactions (via $\mathsf{UserOperation}$). \ding{173}~Bundlers aggregate multiple $\mathsf{UserOperation}$ instances, consolidating them into a single transaction that is then transmitted to the $\mathsf{EntryPoint}$ contract. \ding{174}~Within the $\mathsf{EntryPioint}$ contract, user signatures are verified and transactions initiated by the abstraction layer are processed. \ding{175} The entries logged in $\mathsf{UserOperation}$ prompt the activation of relevant smart contracts, initiating state transitions tailored to their specific prerequisites. \ding{176} Optionally, the user's operations can be aggregated and authenticated using BLS signatures. Meanwhile, transaction fees for user actions are managed by the $\mathsf{Paymaster}$ contract. Consequently, on-chain applications interact with user actions in a manner similar to their interaction with standard external accounts.

\begin{figure}[!htb]
    \centering
	\includegraphics[width=0.99\linewidth]{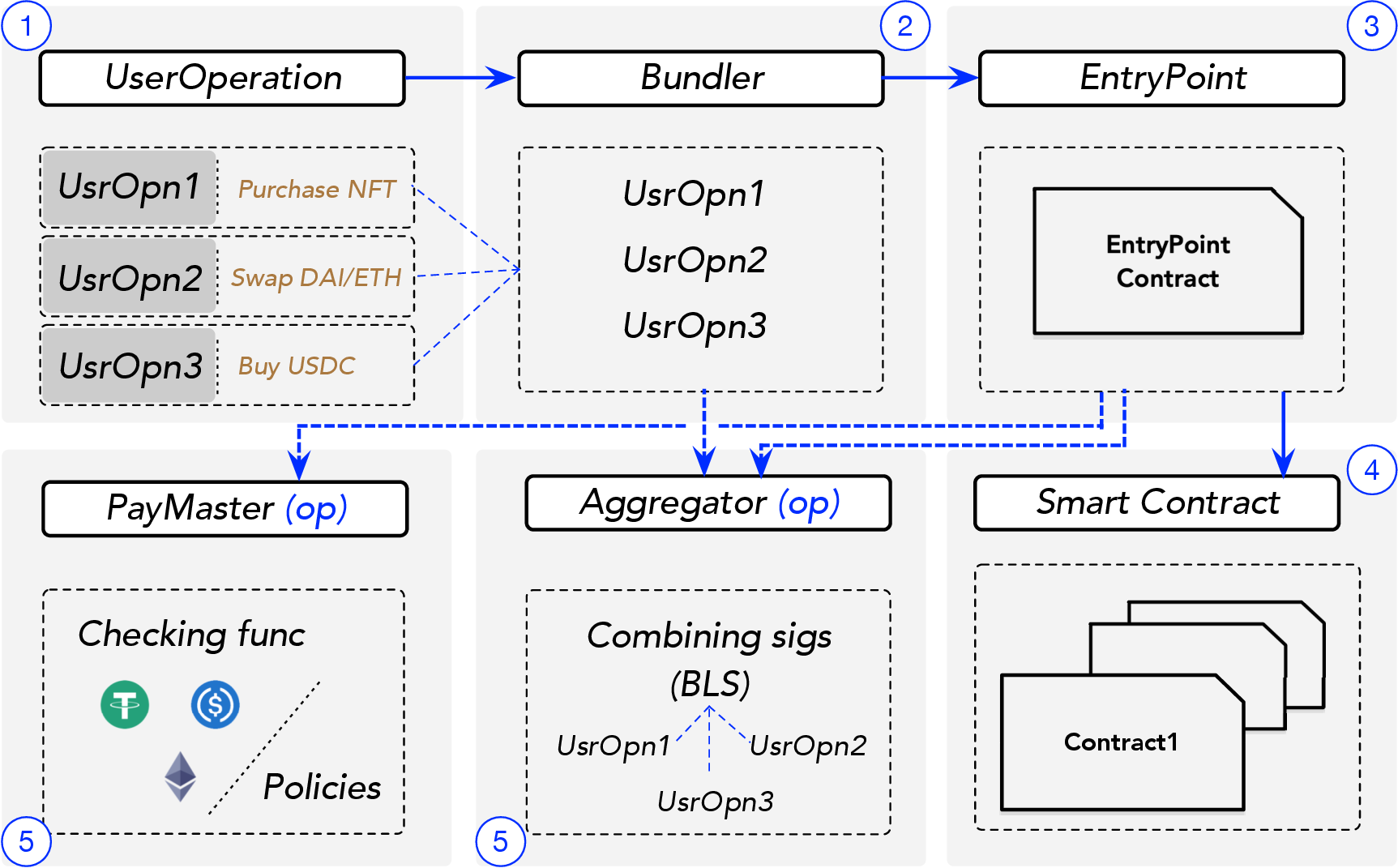}
	\caption{Account abstraction}\label{fig_aa}
 \vspace{-0.4em}
\end{figure}

\noindent\textcolor{teal}{\ding{49} \textit{User operation.}} $\mathsf{UserOperation}$ serves as a transaction-like entity, representing a user's request achieved by transaction events. Within $\mathsf{UserOperation}$, multiple requests and supplementary data can be encapsulated, facilitating the execution of transactions that a CA can perform. While sharing similarities with conventional transactions, $\mathsf{UserOperations}$ exhibit several distinctive characteristics.

\begin{itemize}
\item \textit{Additional fields.} $\mathsf{UserOperation}$ includes new fields in the transaction structure, e.g., $\mathsf{EntryPoint}$, $\mathsf{Bundler}$, $\mathsf{Paymaster}$ and $\mathsf{Aggregator}$. 

\item \textit{Alternate mempool.} $\mathsf{UserOperation}$ is sent to a separate mempool, where bundlers can package them into transactions which get included in a block.

\item \textit{Intent based.} Today, transaction inputs are specific, e.g., swap 2k USDC for 1.2ETH. In contrast, $\mathsf{UserOperation}$ can be decorated with additional metadata to be more outcome-focused, e.g., I want to trade 2k USDC for the most amount of ETH possible.

\end{itemize}

\noindent\textcolor{teal}{\ding{49} \textit{Bundler.}} 
Bundlers constitute a pivotal component of the infrastructure necessary for the realization of EIP-4337. These entities are a specific type of Ethereum node dedicated to facilitating the processing of $\mathsf{UserOperation}$s. These $\mathsf{UserOperation}$s are directed to a network of bundlers, which diligently monitor the mempool. Bundlers efficiently consolidate multiple $\mathsf{UserOperation}$s into a singular transaction, expertly packaging and submitting them to the blockchain on behalf of users. In exchange for this service, bundlers receive compensation for their efforts in executing these tasks.

\smallskip
\noindent\textcolor{teal}{\ding{49} \textit{EntryPoint.}} 
It is a singleton smart contract that undertakes the verification and execution of $\mathsf{UserOperation}$s.

\begin{itemize}
\item \textit{Verification loop.} The process checks the account balance. It involves assessing whether the wallet possesses sufficient funds to cover the maximum potential gas expenditure (derived from the gas field in the UserOp). Transactions lacking adequate funds will be declined.

\item \textit{Execution loop.} Upon verification, the transaction will be executed. Correspondingly, an amount from CA is deducted to reimburse the bundler. This reimbursement is required to cover the gas expenses.
\end{itemize}

\smallskip
\noindent\textcolor{teal}{\ding{49} \textit{Aeggregator (optional).}} 
In scenarios where multiple messages are signed using distinct keys, an aggregator plays a role in producing an authenticated signature. The signature inherently validates the authenticity of all individual signatures within the collection. For accounts associated with particular signature types that enable aggregation (such as BLS), the verification process of the account's signature is deferred to an external contract. This external contract, in turn, is responsible for verifying a solitary signature across an entire bundle, thus streamlining the validation process.

Through the consolidation of multiple signatures into a singular entity, aggregators contribute to the efficient management of data availability. This allows for the validation of numerous bundled $\mathsf{UserOperation}$s in a single consolidated process.

\smallskip
\noindent\textcolor{teal}{\ding{49} \textit{Paymaster (optional).}}
The Paymaster takes charge of implementing gas payment policies, offering developers the capability to offer their end users gas-free interactions through sponsored or ERC-20 gas policies. These policies introduce flexibility in \textit{how gas is paid}, such as the choice of currency (e.g., native ETH or ERC-20 tokens), and \textit{by whom it is paid}. This effectively eliminates the need for users to possess native blockchain tokens to engage with the blockchain.

\subsection{AA Benefits}

Based on its design, AA brings a series of direct benefits that previous CA and EOA do not have.

\begin{itemize}
    \item \textit{Transaction batching.} Capitalizing on the capabilities of the $\mathsf{EntryPoint}$ contract, the consolidation of multiple transactions can guard against frontrunning attacks and potential MEV threats.
    \item \textit{Multisig.} Departing from reliance solely on a single signer  (EOA, private key holder), AA introduces the capability of multi-signatures sourced from a spectrum of entities.  
    \item \textit{Gas abstraction.} ERC-20 tokens are accepted as valid payment for these fees, unshackling the restrictions previously confined to ether (ETH). 
    \item \textit{Social recovery.} AA is engineered to encompass a mechanism for recovery through a social network of trusted associates. This mechanism offers an alternative route to regain access to the wallet.
    \item \textit{Custom module.} Users are empowered with the ability to craft bespoke modules catering to specific functions.
\end{itemize}

\subsection{AA Surroundings}\label{subsec-supports}

\textit{Supportive standards.} 
EIP-\hlhref{https://eips.ethereum.org/EIPS/eip-2771}{2771} is designed to enable gasless transactions for users by introducing the concept of meta-transactions. It permits third parties to cover a user's gas expenses without necessitating modifications to Ethereum. The standards include the following key steps: The transaction signer signs and dispatches transactions to a gas relay. The Gas Relay, operating off-chain, receives these signed requests and subsidizes the gas costs, transforming them into valid transactions that are then routed through a trusted forwarder. This trusted forwarder is essentially a contract entrusted by recipients to accurately verify signatures and nonces.

EIP-\hlhref{https://eips.ethereum.org/EIPS/eip-3047}{3047} is designed to enable users to delegate control of their EOA to a contract, a parallel approach discussed in Sec.\ref{sec-intro}. This feature allows any EOA to function as a smart contract wallet without requiring the deployment of a separate contract. As part of this standard, two new opcodes are introduced to the Ethereum Virtual Machine (EVM): AUTH and AUTHCALL. These opcodes facilitate proper invocation from smart contracts. Although this standard are incorporated into future updates of Ethereum's account infrastructure, it still provides valuable insights.

EIP-\hlhref{https://eips.ethereum.org/EIPS/eip-6900}{6900} introduces a standard aimed at synchronizing the implementation efforts of CA developers and plugin developers. It outlines the creation of a modular CA with the capability to accommodate all plugins adhering to standard protocols. In this standard, CAs are rendered as fully programmable accounts capable of housing assets within on-chain smart contracts. In parallel, account plugins serve as interfaces for smart contracts that facilitate the integration of compositional logic within CAs. This approach enhances data portability for users and alleviates the need for plugin developers to target specific wallet implementations. 

EIP-\hlhref{https://eips.ethereum.org/EIPS/eip-6551}{6551} introduces a token standard primarily focused on NFTs~\cite{wang2021non}. This standard enables the binding of one or multiple Ethereum accounts to an NFT. By doing so, NFTs gain the capability to possess assets and engage with applications, all without necessitating modifications to current smart contracts or infrastructure. These accounts, which are associated with tokens, seamlessly integrate with existing on-chain asset standards and have the potential to encompass forthcoming asset standards. Ethereum's abstract account model offers an enhanced approach to implementing this standard, exemplified by the Web3 game project Sapienz~\cite{sapienz}.
 
\textit{Supportive platforms.} In light of the forthcoming updates to Ethereum, a diverse range of competitive blockchain platforms are also stepping up their game. Notable contenders such as BNB Chain, Avalanche, Optimism, Polygon, StarkNet, and zkSync have all unveiled their respective initiatives for implementing crucial updates.

\textit{Supportive wallet.}  Complementing the infrastructural advancements, an equally pivotal component lies within the realm of wallets. Serving as the hub where users manage, utilize, and safeguard their accounts, wallets play a pivotal role. Multiple wallet service providers, including MetaMash, Argent, Beam, Safe, Trust Wallet, Braavos, and OKX Wallet, have expressed their commitment to seamlessly incorporate these anticipated updates into their products.

\subsection{Real Case Adoption}

\textit{Visa.} As demonstrated in \cite{visa}, Visa is actively exploring a solution that leverages the Paymaster smart contract. This solution aims to abstract the core interactions of blockchain technology and enhance the user payment experience by introducing a self-custodial smart contract wallet. The primary goal is to streamline the process for users conducting transactions within their wallets. One of the intriguing aspects is that users can now use any token to pay for gas fees, and these gas costs will be covered by Paymaster. In more detail, users have the capability to make payments to the Paymaster contract using USDC. Subsequently, the Paymaster contract converts the USDC into ETH and facilitates the transaction on the blockchain network. After a period of on-chain processing, the recipient of the transaction can receive an equivalent amount converted back to USDC. As of now, this particular use case is still in the conceptual verification stage.

\textit{Web3 wallet.} 
By seamlessly integrating AA into existing EOA wallets, the accounts are elevated to the status of smart contract wallets, enriched with programmable logic and functionalities. For instance, Gnosis Safe~\cite{safe} brings forth a \textit{smulti-signature} approach, necessitating multiple authorized entities to provide signatures for the same account, rather than relying solely on individual private keys. Argent~\cite{argent} introduces the concept of \textit{social recovery}, enabling users to recover lost or forgotten private keys. It allows users to utilize email addresses and phone numbers for offline recovery, introducing a familiar two-factor authentication mechanism. Users in Braavos~\cite{bravvos} can access their wallets using the biometric features of their smartphones, such as facial or fingerprint recognition, adding an extra layer of security.

\section{Security Analysis}
\label{sec-security}

We explore the potential security issues that may exist in new account formats after its update. We recall a series of account threats and related smart contract risks and develop a framework to evaluate potential risks in account abstraction.

\smallskip
\noindent\textbf{Focused scope.} Conventional analyses of Ethereum often partition it into four distinct layers (e.g., \cite{chen2020survey}): the application layer (including accounts, smart contracts, and EVM), the data layer (including transactions, blocks, and events), the consensus layer (involving PoX mechanisms and incentives), and the network layer (comprising node discovery, message propagation, and verification). Given the central theme of this endeavor around accounts, we deliberately narrow our purview to the \textit{application layer}, exploring the potential vulnerabilities targeting accounts, smart contracts, and related solidity language. Additionally, our coverage extends to certain elements within the transaction and block layers.

\smallskip
\noindent\textbf{Framework design} (Tab.\ref{tab:vulner}). Inspired by a series of elegant investigations \cite{tsankov2018securify,su2021evil,tolmach2021survey,zou2019smart,krupp2018teether,chen2020survey,brent2020ethainter,wan2021smart}, we distill a succinct overview of vulnerabilities present within the current Ethereum application layer. These vulnerabilities can be broadly attributed to three factors: design pitfalls, vulnerabilities intrinsic to the Solidity programming language, and errors stemming from contract programming. Within each of these categories, we identify and elaborate upon sub-vulnerabilities (\textbf{Vulnerability}) that manifest across diverse functions and processes (\textbf{CauseFmWhere}). Our discussion then delves into the potential of account abstraction to alleviate these vulnerabilities. We examine the feasibility of employing account abstraction to mitigate these issues and explore strategies for implementation that could effectively address these concerns (\textbf{byWhichFunc}).

\begin{table}[!hbt]
\caption{Vulnerabilities in Ethereum application layer}\label{tab:vulner}
\renewcommand\arraystretch{1.1}
\begin{center}
\resizebox{\linewidth}{!}{
\begin{tabular}{l|cc|c} 
        \multicolumn{1}{c}{\rotatebox{90}{\textbf{}}} &  \multicolumn{1}{c}{\cellcolor{gray!10}\textbf{Vulnerability}} & \multicolumn{1}{c}{\cellcolor{gray!10}\textbf{CauseFmWhere}}    & \cellcolor{gray!10}\textbf{byWhichFunc} \\
        
       \cmidrule{1-2}
       \multirow{7}{*}{\rotatebox{90}{\textbf{\makecell{Design}}}} &  \cellcolor{gray!10}Transaction disorder & \cellcolor{gray!10}BlockCreate  &   \cellcolor{gray!10}$\mathsf{Bundler}$ \\ 
       &  \cellcolor{gray!10}Timestamp manipulation & \cellcolor{gray!10}BlockCreate &   \cellcolor{gray!10}$\mathsf{Bundler}$  \\ 
       &   \cellcolor{gray!10}Short address &  \cellcolor{gray!10}InputCheck &   \cellcolor{gray!10}$\mathsf{Bdr}$/$\mathsf{EntryPoint}$    \\ 
       &  \cellcolor{gray!10}Randomness reliance &  \cellcolor{gray!10}BlockCreate &    \cellcolor{gray!10}\xmark    \\ 
       &  \cellcolor{gray!10}Empty account &  \cellcolor{gray!10}StateTrie &  \cellcolor{gray!10}\xmark   \\ 
       &  \cellcolor{gray!10}Under-price opcodes &  \cellcolor{gray!10}GasCost &  \cellcolor{gray!10}\xmark   \\ 
       &  \cellcolor{gray!10}Stack limit &  \cellcolor{gray!10}Execution&    \cellcolor{gray!10}\xmark  \\ 
       \cmidrule{1-2}
       \multirow{6}{*}{\rotatebox{90}{\textbf{\makecell{Solidity}}}}&  \cellcolor{gray!10} Unbounded operations &  \cellcolor{gray!10}GasUsage  & \cellcolor{gray!10}$\mathsf{Paymaster}$   \\ 
       &  \cellcolor{gray!10}Inconsistent call returns &  \cellcolor{gray!10}Exception &  \cellcolor{gray!10}\xmark   \\ 
       &  \cellcolor{gray!10}Pointer unreset & \cellcolor{gray!10}BlankField &  \cellcolor{gray!10}\xmark \\ 
       
       &  \cellcolor{gray!10}Outdated compiler & \cellcolor{gray!10}Compiler &  \cellcolor{gray!10}\xmark   \\ 
       &  \cellcolor{gray!10}Unclear constructor name &  \cellcolor{gray!10}Syntax &    \cellcolor{gray!10}\xmark  \\ 
       &  \cellcolor{gray!10}Type casts &  \cellcolor{gray!10}Design  &  \cellcolor{gray!10}\xmark  \\ 
       
        \cmidrule{1-2}
        
       \multirow{13}{*}{\rotatebox{90}{\textbf{\makecell{Smart Contract}}}} & \cellcolor{gray!10}Reentrancy &\cellcolor{gray!10}Dependence  & \cellcolor{gray!10}$\mathsf{Paymaster}$  \\ 
        & \cellcolor{gray!10}Frozen ether & \cellcolor{gray!10}Dependence &  \cellcolor{gray!10}$\mathsf{Bundler}$    \\ 
        & \cellcolor{gray!10}contract Upgradeability  & \cellcolor{gray!10}Dependence  & \cellcolor{gray!10}\xmark    \\ 
        & \cellcolor{gray!10}Delegatecall injection & \cellcolor{gray!10}Dependence &  \cellcolor{gray!10}\xmark  \\ 
        & \cellcolor{gray!10} Unexpected revert &  \cellcolor{gray!10}Dependence &  \cellcolor{gray!10}\xmark  \\ 
       &  \cellcolor{gray!10}Manipulated balance &  \cellcolor{gray!10}Validation &   \cellcolor{gray!10}$\mathsf{PyMs}$/$\mathsf{UsrOpn}$   \\ 
        & \cellcolor{gray!10}Integer overflow & \cellcolor{gray!10}Validation &    \cellcolor{gray!10}\xmark  \\ 
        &  \cellcolor{gray!10}Insufficient signature & \cellcolor{gray!10}Authentication  & \cellcolor{gray!10}$\mathsf{Bdr}$/$\mathsf{MultiSig}$  \\  
        & \cellcolor{gray!10}Self-destruct contract & \cellcolor{gray!10}Authentication  & \cellcolor{gray!10}$\mathsf{EntryPoint}$    \\
        & \cellcolor{gray!10}$\mathsf{tx.origin}$ & \cellcolor{gray!10}Authentication &    \cellcolor{gray!10}$\mathsf{UsrOpn}$/$\mathsf{Bdr}$  \\ 
       &  \cellcolor{gray!10}Wrong payment & \cellcolor{gray!10}Authentication &  \cellcolor{gray!10}$\mathsf{UsrOpn}$/$\mathsf{Bdr}$ \\ 
       &  \cellcolor{gray!10}Lack of confidentiality &  \cellcolor{gray!10}Authentication &  \cellcolor{gray!10}\xmark   \\ 
       &  \cellcolor{gray!10}Erroneous visibility & \cellcolor{gray!10}Authentication &  \cellcolor{gray!10}\xmark   \\

       \cmidrule{3-4}
       \multicolumn{3}{c}{} &   \multicolumn{1}{c}{\cellcolor{gray!10}\textit{\textbf{AA updates}}}   \\ 
\end{tabular}
}
    \begin{tablenotes}
      \footnotesize
      \item[] \quad  \textbf{Abbr.} PyMs - Paymaster; UsrOpn - UserOperation; Bdr - Bundler.
     \end{tablenotes}
\end{center}
\vspace{-0.4em}
\end{table}

\smallskip
\noindent\textbf{Analyses.} Our discussions are categorized into two facets, contingent upon the effectiveness of AA updates.

\smallskip
\noindent\textit{\cmark\, Transaction disorder.} This vulnerability centers around a concurrency issue where the blockchain's future state is contingent upon the sequence of transaction execution, a process influenced by miners who group transactions into blocks based on incentives. The design of AA effectively mitigates this problem by employing bundlers to manage and submit transactions. Unlike miners or validators, bundlers have a limited scope of impacting transaction orders, as they handle a relatively small number of transactions.

\smallskip
\noindent\textit{\cmark\, Timestamp manipulation.} This vulnerability happens when a contract utilizes $\mathsf{block.timestamp}$ in crucial operations or as a means of generating randomness. It can be exploited by a malicious miner who has the capability to manipulate the value of $\mathsf{block.timestamp}$ (evidenced by~\cite{zhang2023time}). The introduction of account abstraction helps address this concern, as AA reduces reliance on $\mathsf{block.timestamp}$ and shifts focus to bundlers, which are better suited for managing transactions and ensuring more controlled and secure interactions.

\smallskip
\noindent\textit{\cmark\, Short address.} The vulnerability stems from the EVM's absence of address validity checking. In the process of encoding during contract invocation, if the encoded arguments' length is insufficient, the EVM compensates by adding extra zeros to ensure 32 bytes. The AA design, through the involvement of the $\mathsf{bundler}$ and $\mathsf{EntryPoint}$ contract, can mitigate this issue. The design can effectively check the length of Ethereum addresses via $\mathsf{msg.data}$.

\smallskip
\noindent\textit{\xmark\, Randomness reliance.} This vulnerability revolves around the manipulation of seed (a form of randomness) by malicious miners in gambling and lottery contracts that use pseudorandom number generation. AA might not directly enhance the security of random seed creation. This is because the process of generating random seeds typically involves functions within the contract logic rather than being directly embedded in accounts, which is the primary focus of AA.

\smallskip
\noindent\textit{\xmark\, Empty account.}
Malicious actors leverage empty accounts (devoid of nonce, balance, code, and storage) to initiate a DoS attack. These empty accounts still necessitate tracking within the Ethereum state trie and result in extended transaction processing times. AA is not equipped to mitigate this vulnerability, as it currently lacks verification mechanisms for empty accounts. However, the issue has been addressed through the implementation of EIP-\hlhref{https://eips.ethereum.org/EIPS/eip-161}{161}.

\smallskip
\noindent\textit{\xmark\, Under-price opcodes.} Ethereum uses the gas mechanism to deter abuse of computing resources, but this vulnerability emerges when the gas cost for resource consumption is improperly set. The exploitation of under-priced opcodes will consume a disproportionate amount of computing resources.  This issue is not specifically addressed by the design of account abstraction, as it pertains to the gas pricing and resource allocation mechanisms within Ethereum.

\smallskip
\noindent\textit{\xmark\, Stack limit.} Solidity would trigger an exception and terminate the call due to its rigid limit of 1,024 frames in the EVM call stack.  Malicious actors could recursively call a contract, causing it to reach the maximum stack depth and fail subsequent external calls. While account abstraction cannot inherently address this specific vulnerability, changes in the Ethereum protocol (e.g.,  EIP-\hlhref{https://eips.ethereum.org/EIPS/eip-150}{150}) can indirectly impact AA as well as other aspects of the Ethereum ecosystem.

\smallskip
\noindent\textit{\cmark\, Unbounded operations.} The vulnerability arises when a contract's execution demands a greater amount of gas than the permitted block gas limit. This occurs due to the inclusion of unbounded operations, such as loops, within the contract. AA design offers effective solutions to counter such issues in two primary ways: Firstly, the paymaster contract defines a fee policy that establishes clear limitations on gas limits. This mechanism ensures that transactions adhere to predefined gas constraints, preventing situations where execution demands exceed the allowable limit. Secondly, after AA's update, every loop in smart contracts is invoked by an internal account rather than an external account. This setup grants the contract owner the ability to terminate loops if necessary, enhancing control over potential gas consumption issues.

\smallskip
\noindent\textit{\xmark\, Inconsistent call returns.} This vulnerability stems from the inconsistency between the two return values from different calling routes: (i) directly referencing the callee contract instance and (ii) using low-level methods like send and call. Route-(i) will return the call back to the caller while Route-(ii) returns false to the caller. The inconsistency in behavior persists in the current state of the Solidity compiler, lacking rectification. AA design does not inherently address this issue, but updates or changes in the Solidity compiler or EVM could potentially influence AA as well.

\smallskip
\noindent\textit{\xmark\, Pointer unreset.} The vulnerability arises from Solidity's behavior regarding uninitialized compound local variables, which, when not explicitly initialized, leads to the reference defaulting to slot 0 and overwriting content from that point onwards. AA design does not address this particular language-level issue, but it's fortunate that the vulnerability was rectified by updating the version of Solidity.

\smallskip
\noindent\textit{\xmark\, Outdated compiler.} This vulnerability surfaces when a contract is compiled using an outdated compiler that harbors bugs. While AA design may not directly address this concern, it aligns with the solution: opting for an up-to-date compiler can substantially reduce the associated risks. Keeping the compiler current is a key practice to ensure the solidity and security of the compiled contracts.

\smallskip
\noindent\textit{\cmark\, Unclear constructor name.}  The vulnerability stems from the misnaming of the constructor function, inadvertently allowing unauthorized parties to take ownership of the contract. While AA design cannot directly solve this issue, it is fortunate that the problem was mitigated through the introduction of the \textit{constructor} keyword

\smallskip
\noindent\textit{\xmark\, Type casts.} A contract can invoke another contract's function by directly referencing the callee contract's instance. However, the verification process carried out by the Solidity compiler only assesses whether the invoked function has been declared, and it doesn't extend to verifying child functions that the invoked function inherits or parent functions that trigger it. This loophole can lead to unintended function execution. AA design does not offer a solution for this vulnerability, as it stems from Solidity's insufficient type system. In this context, AA's mechanisms don't directly address this issue.

\smallskip
\noindent\textit{\cmark\, Reentrancy.} The vulnerability arises when an external contract initiates a function within the caller contract before the latter's ongoing execution concludes—essentially manifesting as a cyclic call. This exploitable invocation persists until the caller contract depletes its gas resources. The foundation of this vulnerability can be attributed to two pivotal factors: (i) a contract's decision-making hinges on specific state variables, which necessitate updating prior to invoking another contract; and (ii) the absence of a predefined gas limit for the transition. The introduction of AA can offer a partial solution to this issue. AA's advantages, including reduced gas consumption and the ability to control transactions precisely through the paymaster contract, contribute to its potential mitigation. Additionally, user operations are treated as distinct inputs through independent transactions, fostering an environment free from dependencies on other contracts. This segregation of operations enhances security by reducing the risk of vulnerability to such cyclic call-related attacks.

\smallskip
\noindent\textit{\cmark\, Frozen ether.} The vulnerability arises from a scenario in which users can deposit funds into their contract accounts but are unable to later access or spend those deposited funds. Two primary factors contribute to this vulnerability: (i) contracts not furnishing a function for the expenditure of funds, and (ii) the inadvertent or deliberate termination of the callee contract. Account abstraction can offer partial mitigation of this vulnerability due to its transaction batching made by bundlers. With each user-requested operation, such as depositing funds into a smart contract, the transaction bundler gains an enhanced capability to inspect the contract before sending tokens. This heightened level of examination surpasses that of a standard user. The transaction bundler possesses the potential to ensure that operations transpire within the confines of the originating contract's security.

\smallskip
\noindent\textit{\xmark\, Upgraded contract.} Contract upgrading can be approached through two methods: (i) Dividing a contract into a proxy contract (non-updatable) and a logic contract (updatable). (ii) Employing a registry contract to maintain a record of the updated contracts. While the design of account abstraction is separate from these complexities, its contract functionalities align with its principles of upgradability.

\smallskip
\noindent\textit{\xmark\, Delegatecall injection.} EVM offers an opcode called $\mathsf{delegatecall}$, allowing the bytecode of a callee contract to be inserted into the bytecode of the caller contract. The vulnerability arises due to the possibility that a callee contract can modify state variables within the caller contract. Account abstraction does not inherently provide a distinct method to mitigate this particular vulnerability. The effectiveness of AA's security in mitigating the issue hinges on its inherent design and solution. Fortunately, addressing this can be achieved via a straightforward means by enhancing its declaration.

\smallskip
\noindent\textit{\xmark\, Unexpected revert.} This vulnerability stems from the situation where a callee contract causes the execution of a caller contract to be reverted. The design of account abstraction does not offer a significantly improved mitigation for this issue, as it fundamentally stems from contract programming practices. Consequently, this vulnerability might still be pertinent even within AA's contract functionalities.

\smallskip
\noindent\textit{\cmark\, Manipulated balance.} This vulnerability surfaces when a contract's decision-making relies on the values of $\mathsf{this.balance}$ or $\mathsf{address(this).balance}$. These values can be manipulated by an attacker to gain unauthorized access to funds. The design update introduced by AA can potentially influence this vulnerability. In AA, all account addresses are explicitly bound with balances, unlike the implicit nature of the former design. This alteration could impact the dynamics of this vulnerability.

\smallskip
\noindent\textit{\xmark\, Integer overflow.} This vulnerability arises when the result of an arithmetic operation surpasses the range of a Solidity data type. Regrettably, neither the Solidity compiler nor the EVM incorporates mechanisms to identify integer overflow or underflow. Account abstraction design limitations prevent it from overcoming these challenges. However, a solution to mitigate this vulnerability is within reach through the utilization of the SafeMath library, which has already proven effective in handling such arithmetic issues.

\smallskip
\noindent\textit{\cmark\, Insufficient signature.} The vulnerability surfaces when a sender channels funds to multiple recipients via a proxy contract, bypassing individual transactions. In this mechanism, the proxy contract evaluates the authenticity of digital signatures from senders. If these signatures lack essential data (such as nonce or proxy contract address), a malicious recipient can exploit the situation to replay the message repeatedly, facilitating the withdrawal of surplus payments. In AA's design, the role of the bundler mirrors that of a proxy agent. The bundler manages the accumulation of multiple transactions, functioning as a trusted entity. This approach not only aligns with the principles of a proxy contract but also serves as a safeguard against replay attacks. Further, the adoption of multi-sig techniques in AA can also contribute to the mitigation of issues.

\smallskip
\noindent\textit{\cmark\, Self-destruct contract.} When the owner of a contract or a third party utilizes the self-destruct method to terminate a contract, the contract's bytecode and storage will be erased. The vulnerability emerges from inadequate authentication mechanisms embedded within the contract. AA can indeed mitigate this issue by introducing multi-factor authentication, a more robust approach. This mandates the approval of multiple parties before a suicide operation is executed. By implementing this safeguard, AA enhances security by fortifying the authentication process and reducing the likelihood of unauthorized termination.

\smallskip
\noindent\textit{\cmark\, $\mathsf{tx.origin}$.} The problem originates from the utilization of $\mathsf{tx.origin}$, a global variable within Solidity that identifies the original EOA responsible for triggering the transaction. The vulnerability arises when a contract employs $\mathsf{tx.origin}$ for authorization, thereby exposing it to phishing attacks. AA addresses this concern through its integration of EOA with contract invocation. The integration is pivotal as AA replaces $\mathsf{tx.origin}$ with $\mathsf{msg.sender}$ in default authentication. This shift is now a necessary step to bolster security against vulnerabilities stemming from $\mathsf{tx.origin}$ usage.

\smallskip
\noindent\textit{\cmark\, Wrong payment.} This security issue arises due to the absence of identity verification when a caller invokes a function to transfer Ether to a random address. As seen in the mitigation strategies employed for previous balance-related vulnerabilities, the authentication mechanisms and the role of the bundler within account abstraction can aid in alleviating this problem. This ensures that only authorized entities are permitted to initiate such actions, thwarting unauthorized withdrawals.

\smallskip
\noindent\textit{\xmark\, Lack of confidentiality.} Transaction details within a blockchain are inherently public due to the nature of the technology. While designating a state variable as private can limit other contracts' access, the value of such a variable can still be inferred from transaction data. It's worth noting that AA design does not specifically address the intricacies of transaction confidentiality; however, it aligns with the overall progression of smart contract evolution. Mitigation strategies encompass the implementation of cryptographic techniques like commitment schemes~\cite{wang2020preserving} and zero-knowledge proofs~\cite{kosba2016hawk}. Additionally, hardware-based solutions~\cite{li2022sok} could also be explored to enhance the security of sensitive data within an untrusted environment.

\smallskip
\noindent\textit{\xmark\, Erroneous visibility.} This vulnerability stems from inaccurately defining the visibility of a function, where the default public visibility can be exploited by attackers to improperly access functions. AA design doesn't directly address this issue but aligns with its solution. This vulnerability is mitigated by requiring explicit specification of function visibility, a measure AA's design also adheres to.

\section{Further Discussions}
\label{sec-discu}

\subsection{More about AA Design}

\textit{AA represents an application-level update with its primary impact centered around the application layer.} Aligned with the idea outlined in EIP-\hlhref{https://eips.ethereum.org/EIPS/eip-4337}{4337}, AA introduces perfect backward compatibility, minimizing its effects on consensus layers and underlying structures. The fact is evident in security enhancements, which effectively tackle many challenges originating from the application layer rather than fundamental design flaws (refer to Tab.\ref{tab:vulner}).

\textit{Rather than directly merging EOA and CA, AA incorporates key functionalities and delegates subsequent tasks to related contracts in tandem.} Notably, separating the functions required to be executed within the smart contract and allocating a portion of them to the user's account enhances overall efficiency. The corresponding batch processing isn't limited to user operations but also extends to fee payments. 

\textit{The user experience is markedly improved when employing an AA account as opposed to an EOA.} While a series of interconnected contracts work harmoniously to support the functionality of an AA account, users perceive it as effortlessly accessible as popular Web2 Apps (e.g., shopping, and banking). Complex technical jargon is relegated to the backend, and this will spare regular users from unnecessary cognitive load. The user interface offers a friendly entry point.

\subsection{AA Opportunities}

\textit{Ease accessibility to Web3.} Account abstraction not only streamlines the interaction with DApps but also significantly enhances the user experience within Web3.  By embracing AA, users gain the ability to tailor their accounts to operate exclusively under specific conditions, which significantly differs from previous solutions that rely on complicated invocation across functions or contracts~\cite{yu2022towards}. Users are empowered with a greater degree of control over their customized programming functions. For instance, unlike the traditional multi-sig setup, where the conditions for executing transactions are relatively standardized, account abstraction allows for a more personalized set of conditions. This adaptability ensures that the operations are only executed when predetermined criteria are met, such as a predefined number of signatories.

\textit{Promoting Layer-2 (L2).}  The fundamental L2 projects revolve around alleviating the substantial computational load on L1 chains. As a result, a significant focus of various L2 solutions lies in streamlining the intricate processing logic inherent in DApps. This optimization not only reduces the overall gas cost but also enriches the user experience through the introduction of extended functionalities. To realize this goal, numerous L2 projects have embraced the notion of incorporating account abstraction, marking a pivotal step toward achieving their mission. Notable examples~\cite{binanceaa} include zkSync, which seamlessly integrates the IAccount interface, and StarkNet, with the well-regarded platform Argent. Similarly, Optimistic rollups also witness the integration of customized APIs.

\section{Conclusion}
\label{sec-conclusion}

In this paper, we explore the notion of account abstraction (featured by EIP-\hlhref{https://eips.ethereum.org/EIPS/eip-4337}{4337}), which is formally involved the sixth stage of Ethereum's roadmap. We study its operating mechanism, key features, and surrounding developments. We also examine its security by assessing a set of related criteria. Our results reveal the scope and extent of the security improvements introduced by the adoption of account abstraction. To our knowledge, this work provides the first formal AA study.

\newpage
{\footnotesize \bibliographystyle{IEEEtran}
\bibliography{bib}}

\begin{thebibliography}{10}
\providecommand{\url}[1]{#1}
\csname url@samestyle\endcsname
\providecommand{\newblock}{\relax}
\providecommand{\bibinfo}[2]{#2}
\providecommand{\BIBentrySTDinterwordspacing}{\spaceskip=0pt\relax}
\providecommand{\BIBentryALTinterwordstretchfactor}{4}
\providecommand{\BIBentryALTinterwordspacing}{\spaceskip=\fontdimen2\font plus
\BIBentryALTinterwordstretchfactor\fontdimen3\font minus
  \fontdimen4\font\relax}
\providecommand{\BIBforeignlanguage}[2]{{%
\expandafter\ifx\csname l@#1\endcsname\relax
\typeout{** WARNING: IEEEtran.bst: No hyphenation pattern has been}%
\typeout{** loaded for the language `#1'. Using the pattern for}%
\typeout{** the default language instead.}%
\else
\language=\csname l@#1\endcsname
\fi
#2}}
\providecommand{\BIBdecl}{\relax}
\BIBdecl

\bibitem{wood2014ethereum}
G.~Wood \emph{et~al.}, ``Ethereum: A secure decentralised generalised
  transaction ledger,'' \emph{Ethereum project yellow paper}, vol. 151, no.
  2014, pp. 1--32, 2014.

\bibitem{account}
@jmcook1186, ``Ethereum accounts,'' \emph{Retrived
  \url{https://ethereum.org/en/developers/docs/accounts/}}, 2023.

\bibitem{erc4337}
``Ethereum: {ERC}-4337,'' \emph{\url{https://www.erc4337.io/}}, 2023.

\bibitem{ethereumaa}
Ethereum, ``Ethereum roadmap: Account abstraction,''
  \emph{\url{https://ethereum.org/en/roadmap/account-abstraction/}}, 2023.

\bibitem{evm}
T.~T., ``Ethereum {EVM} illustrated,''
  \emph{\url{https://takenobu-hs.github.io/downloads/ethereum_evm_illustrated.pdf}},
  2023.

\bibitem{singh2023account}
A.~K. Singh, I.~U. Hassan, G.~Kaur, S.~Kumar \emph{et~al.}, ``Account
  abstraction via singleton entrypoint contract and verifying paymaster,''
  \emph{International Conference on Edge Computing and Applications (ICECAA)},
  pp. 1598--1605, 2023.

\bibitem{binanceaa}
C.~Colin, ``Biance research: A primer on account abstraction,''
  \emph{\url{https://research.binance.com/static/pdf/a-primer-on-account-abstraction.pdf}},
  2023.

\bibitem{metamaskaa}
A.~Emmanuel, ``Account abstraction: Past, present, future,''
  \emph{\url{https://metamask.io/news/latest/account-abstraction-past-present-future/}},
  2023.

\bibitem{Alchemyaa}
Alchemy, ``How {ERC-4337} supports account abstraction,''
  \emph{\url{https://www.alchemy.com/learn/account-abstraction}}, 2023.

\bibitem{coinmarketcapaa}
Moderate, ``Coinmarketcap: Account abstraction,''
  \emph{\url{https://coinmarketcap.com/alexandria/glossary/account-abstraction}},
  2023.

\bibitem{Hackernoonaa}
aleksandrmalyshev, ``What is account abstraction and why is everyone talking
  about it?''
  \emph{\url{https://hackernoon.com/what-is-account-abstraction-and-why-is-everyone-talking-about-it}},
  2023.

\bibitem{wang2022exploring}
Q.~Wang, R.~Li, Q.~Wang, S.~Chen, M.~Ryan, and T.~Hardjono, ``Exploring {Web3}
  from the view of blockchain,'' \emph{arXiv preprint arXiv:2206.08821}, 2022.

\bibitem{park2023beyond}
S.~Park, J.~H. Lee, S.~Lee, J.~H. Chun, H.~Cho, M.~Kim, H.~K. Cho, and S.-M.
  Moon, ``Beyond the blockchain address: Zero-knowledge address abstraction,''
  \emph{Cryptology ePrint Archive}, 2023.

\bibitem{wang2021non}
Q.~Wang, R.~Li, Q.~Wang, and S.~Chen, ``Non-fungible token ({NFT}): Overview,
  evaluation, opportunities and challenges,'' \emph{arXiv preprint
  arXiv:2105.07447}, 2021.

\bibitem{sapienz}
``Sapienz wallet,'' \emph{\url{https://www.sapienz.xyz/}}, 2023.

\bibitem{visa}
B.~Andrew, G.~Catherine, R.~Srini, M.~Mohsen, and K.~Ranjit, ``Visa: Auto
  payments for self-custodial wallets,''
  \emph{\url{https://usa.visa.com/solutions/crypto/auto-payments-for-self-custodial-wallets.html}},
  2023.

\bibitem{safe}
``Safe wallet,'' \emph{\url{https://safe.global/wallet}}, 2023.

\bibitem{argent}
``Argent wallet,'' \emph{\url{https://www.argent.xyz/}}, 2023.

\bibitem{bravvos}
``Bravvos wallet,'' \emph{\url{https://braavos.app/}}, 2023.

\bibitem{chen2020survey}
H.~Chen, M.~Pendleton, L.~Njilla, and S.~Xu, ``A survey on {Ethereum systems}
  security: Vulnerabilities, attacks, and defenses,'' \emph{ACM Computing
  Surveys (CSUR)}, vol.~53, no.~3, pp. 1--43, 2020.

\bibitem{tsankov2018securify}
P.~Tsankov, A.~Dan, D.~Drachsler-Cohen, A.~Gervais, F.~Buenzli, and M.~Vechev,
  ``Securify: Practical security analysis of smart contracts,'' in
  \emph{Proceedings of the ACM SIGSAC Conference on Computer and Communications
  Security (CCS)}, 2018, pp. 67--82.

\bibitem{su2021evil}
L.~Su, X.~Shen, X.~Du, X.~Liao, X.~Wang, L.~Xing, and B.~Liu, ``Evil under the
  sun: Understanding and discovering attacks on ethereum decentralized
  applications,'' in \emph{USENIX Security Symposium (USENIX Sec)}, 2021, pp.
  1307--1324.

\bibitem{tolmach2021survey}
P.~Tolmach, Y.~Li, S.-W. Lin, Y.~Liu, and Z.~Li, ``A survey of smart contract
  formal specification and verification,'' \emph{ACM Computing Surveys (CSUR)},
  vol.~54, no.~7, pp. 1--38, 2021.

\bibitem{zou2019smart}
W.~Zou, D.~Lo, P.~S. Kochhar, X.-B.~D. Le, X.~Xia, Y.~Feng, Z.~Chen, and B.~Xu,
  ``Smart contract development: Challenges and opportunities,'' \emph{IEEE
  Transactions on Software Engineering (TSE)}, vol.~47, no.~10, pp. 2084--2106,
  2019.

\bibitem{krupp2018teether}
J.~Krupp and C.~Rossow, ``{teEther}: Gnawing at {Ethereum} to automatically
  exploit smart contracts,'' in \emph{USENIX Security Symposium (USENIX Sec)},
  2018, pp. 1317--1333.

\bibitem{brent2020ethainter}
L.~Brent, N.~Grech, S.~Lagouvardos, B.~Scholz, and Y.~Smaragdakis, ``Ethainter:
  a smart contract security analyzer for composite vulnerabilities,'' in
  \emph{Proceedings of the ACM SIGPLAN Conference on Programming Language
  Design and Implementation (SIGPLAN)}, 2020, pp. 454--469.

\bibitem{wan2021smart}
Z.~Wan, X.~Xia, D.~Lo, J.~Chen, X.~Luo, and X.~Yang, ``Smart contract security:
  A practitioners' perspective,'' in \emph{IEEE/ACM International Conference on
  Software Engineering (ICSE)}.\hskip 1em plus 0.5em minus 0.4em\relax IEEE,
  2021, pp. 1410--1422.

\bibitem{zhang2023time}
X.~Zhang \emph{et~al.}, ``Time-manipulation attack: Breaking fairness against
  proof of authority {Aura},'' in \emph{Proceedings of the ACM Web Conference
  (WWW)}, 2023, pp. 2076--2086.

\bibitem{wang2020preserving}
Q.~Wang, B.~Qin, J.~Hu, and F.~Xiao, ``Preserving transaction privacy in
  {Bitcoin},'' \emph{Future Generation Computer Systems (FGCS)}, vol. 107, pp.
  793--804, 2020.

\bibitem{kosba2016hawk}
A.~Kosba, A.~Miller, E.~Shi, Z.~Wen, and C.~Papamanthou, ``Hawk: The blockchain
  model of cryptography and privacy-preserving smart contracts,'' in \emph{IEEE
  Symposium on Security and Privacy (SP)}.\hskip 1em plus 0.5em minus
  0.4em\relax IEEE, 2016, pp. 839--858.

\bibitem{li2022sok}
R.~Li \emph{et~al.}, ``{SoK}: {TEE}-assisted confidential smart contract,''
  \emph{Proceedings on Privacy Enhancing Technologies}, vol.~3, pp. 711--731,
  2022.

\bibitem{yu2022towards}
G.~Yu, X.~Wang \emph{et~al.}, ``Towards web3 applications: Easing the access
  and transition,'' \emph{arXiv preprint arXiv:2210.05903}, 2022.

\end{thebibliography}

\end{document}